\documentclass[conference]{IEEEtran}
\IEEEoverridecommandlockouts
\usepackage{cite}
\usepackage{amsmath,amssymb,amsfonts}
\usepackage{algorithmic}
\usepackage{graphicx}
\usepackage{textcomp}
\usepackage{enumitem}
\usepackage{xcolor}
\usepackage{algorithm}
\usepackage{subfig}
\def\BibTeX{{\rm B\kern-.05em{\sc i\kern-.025em b}\kern-.08em
    T\kern-.1667em\lower.7ex\hbox{E}\kern-.125emX}}

\renewcommand{\baselinestretch}{0.84}
\begin{document}

\title{Hierarchical Cooperative MARL for Joint Downlink PRB and Power Allocation in a 5G System\\
\thanks{This material is based upon work supported by the National Science Foundation under Grant Numbers CNS-2202972, CNS-2318726, and CNS-2232048.}
}
\author{
\IEEEauthorblockN{
Alireza Ebrahimi Dorcheh,
Tolunay Seyfi,
Ryan Barker,
Fatemeh Afghah
}
\IEEEauthorblockA{
\textit{Holcombe Department of Electrical and Computer Engineering} \\
Clemson University, USA \\
\{ alireze, tseyfi, rcbarke, fafghah\}@clemson.edu
}
}
\maketitle

\begin{abstract}
 
Efficient 5G downlink radio resource management requires jointly optimizing user scheduling and transmit-power allocation under time-varying wireless conditions. This is challenging in OFDMA because PRB assignment is combinatorial, power allocation is continuous, and performance depends on channel evolution, link adaptation, and long-term fairness. We propose a hierarchical cooperative multi-agent reinforcement learning framework with staged curriculum training for joint downlink PRB and power allocation in a physically grounded 5G environment. Simulations use Sionna for system-level modeling and Sionna RT for wireless scene construction and mobility-aware ray-traced channels. The task is decomposed into two stages: a PRB agent learns user-level resource shares, mapped to exact PRB assignments by a deterministic channel-aware quota resolver, and a power agent allocates base-station power across users and assigned PRB-symbol resources. The framework runs in a cross-layer loop with adaptive modulation and coding, HARQ feedback, outer-loop link adaptation, and a fairness-aware reward using smoothed throughput and Jain's fairness index. A three-phase curriculum improves stability by training PRB allocation, power control, and joint fine-tuning. Under matched channel realizations, comparisons with an equal-power PF scheduler and two ablations isolating the learned PRB and power-control components show both components improve throughput distribution over PF, while the full PRB and power controller achieves the largest cell-throughput gain with only a modest reduction in Jain's fairness index.
\end{abstract}


\section{Introduction}

Downlink radio resource management remains central in 5G New Radio (NR). The next-generation NodeB (gNB) schedules physical resource blocks (PRBs), each spanning 12 subcarriers within a bandwidth part, while physical downlink shared channel (PDSCH) resource allocation and modulation-and-coding-scheme (MCS) selection follow standardized procedures \cite{3gpp38211,3gpp38214}. The gNB must jointly select users, assign PRBs, and allocate transmit power under base-station (BS) power constraints \cite{3gpp38104}. These decisions are coupled: favorable PRBs can be wasted by poor power allocation, whereas aggressive power concentration can raise instantaneous rate but reduce fairness. The challenge further increases with channel variation, mobility, hybrid automatic repeat request (HARQ) feedback, and link adaptation \cite{3gpp38321,3gpp38214}. Proportional-fair (PF) scheduling is an interpretable baseline for balancing channel quality and long-term service, but it is hand-crafted and often paired with fixed or equal-power transmission \cite{kelly1998rate,jalali2000data,kushner2004convergence}. Joint PRB/subcarrier and power allocation has been studied through optimization and learning \cite{setayesh2020joint,cheng2023joint,elsayed2019reinforcement,li2023joint,kim2025joint}. Yet many methods cast the task as one large decision, causing high-dimensional actions, unstable training, limited interpretability, and reliance on simplified channels that miss realistic geometry and dynamics. We address these limitations with a hierarchical cooperative multi-agent reinforcement learning (MARL) framework built on Sionna system-level simulation and Sionna RT scene-aware channels. The PRB agent learns user-level shares instead of a full PRB map; a deterministic channel-aware quota resolver converts these shares into exact assignments. Conditioned on this schedule, a factorized power agent allocates each user's total power and shapes it over assigned PRB-symbol resources. This keeps actions compact while preserving scheduling--power coupling. A key strength of the proposed framework is that it is evaluated in a physically grounded street-canyon scenario with mobile users and Sionna RT ray traced propagation. Its cross layer loop links scheduling and power decisions to signal-to-interference-plus-noise ratio (SINR), adaptive MCS, HARQ, and long term throughput fairness. The reward combines cell throughput and Jain's fairness index. Empirically, the learned PRB and power control components each improve efficiency over PF with equal power, and their joint use provides the largest throughput gain with modest fairness reduction. To improve optimization stability, we adopt a staged curriculum-based training procedure. The PRB policy is trained first, the power policy is then trained on top of the learned scheduling behavior, and both policies are finally fine-tuned jointly. This training strategy aligns naturally with the hierarchical structure of the control problem and reduces the difficulty of simultaneous exploration over scheduling and power-control behavior.

The main contributions of this paper are summarized as follows:

\begin{itemize}[leftmargin=*]
    \item We formulate joint downlink PRB/power allocation as hierarchical cooperative MARL with sequential PRB and power agents optimizing a common throughput--fairness objective.
    \item We implement the system-level simulator in Sionna with Sionna RT for physically grounded, mobility-aware, ray-traced channels.
    \item We introduce a deterministic quota-based PRB resolver that maps learned user shares to feasible channel-aware PRB assignments.
    \item We design a factorized power policy for inter-user power partitioning and intra-user PRB-symbol power shaping.
    \item We use staged curriculum training and matched-channel ablations for stable learning and fair comparison with PF, PRB-only, and power-only controllers.
\end{itemize}

%The rest of the paper is organized as follows. Section \ref{sysModel} presents the system model and problem formulation. Section \ref{proposedMethod} describes the proposed hierarchical cooperative MARL framework and the staged training procedure. Section \ref{simulation} details the Sionna-based simulation setup and evaluation methodology. Section \ref{results} reports numerical results and discussion, and Section \ref{conclusion} concludes the paper.
%% COMPRESSED VERSION
The remainder presents the system model, proposed framework, simulation methodology, results, and conclusion.

\section{Related Work}

\subsection{Joint Optimization of Radio Resources}

Joint radio resource optimization is well studied when frequency assignment and power control interact. For example, \cite{setayesh2020joint} formulates joint PRB/power allocation for eMBB/URLLC coexistence in 5G C-RAN, and \cite{cheng2023joint} extends this direction to 5G H-CRAN with cross-layer interference and energy-efficiency objectives. These works show that PRB/RB assignment and power allocation should not be optimized independently because frequency gains can be lost under poor power splits, while aggressive power concentration can reduce fairness.

Fairness-aware resource optimization has also been studied in heterogeneous systems. \cite{jang2025joint} considers joint user association and resource allocation with multi-level fairness, showing that differentiated fairness can be built directly into distributed radio-resource optimization. Although not downlink PRB scheduling with symbol-level power shaping, it supports our use of an explicit throughput--fairness objective. Reinforcement learning is attractive for high-dimensional or non-convex online allocation. Prior work applies RL to joint URLLC power/resource allocation~\cite{elsayed2019reinforcement}, OFDM subcarrier/power allocation~\cite{li2023joint}, decomposed IAB scheduling/resource allocation~\cite{kim2025joint}, and O-RAN slice-level PRB allocation, where DORA uses PPO with deterministic intra-slice scheduling to reduce complexity~\cite{DORA}. Together, these works motivate structured learning-based control when monolithic action spaces are impractical.

Taken together, these studies strongly motivate our formulation: joint scheduling and power control matter, fairness-aware objectives matter, and decomposition is often necessary for tractable learning. However, most prior works in this stream either remain optimization-driven, operate at subcarrier or association level rather than downlink PRB-level scheduling with intra-user power shaping, or do not explicitly consider a sequential hierarchical policy in which user-level PRB shares are first inferred and then resolved into exact PRB assignments before power is allocated.

\subsection{Digital-Twin-Based and Physically Grounded Resource Management} Digital twins (DTs) and physically grounded wireless environments are increasingly used for radio access network (RAN) control control because scheduling and power decisions depend on geometry, mobility, propagation, and protocol-state evolution, which abstract channels often oversimplify. For example, \cite{tarafder2026digital} develops a DT-enabled framework for site-specific radio resource management in a NextG aerial corridor, combining high-fidelity ray tracing with deep reinforcement learning (DRL) for BS association and beam selection. \cite{elloumi2026uplink} studies uplink RB scheduling and power allocation in a DT-integrated open radio access network (O-RAN) Internet-of-Drones network with geo-referenced 3D DTs, GPU-accelerated propagation, and RL. These works reinforce the value of realistic, site-specific evaluation for resource-block and power decisions. DT-assisted learning also appears in edge offloading and synchronization: \cite{chen2023digital} uses DT-assisted RL for microservice offloading and bandwidth allocation, while \cite{tong2025continual} studies continual RL for DT synchronization through resource-block allocation and device scheduling. Although outside 5G downlink PHY/MAC PRB-power scheduling, they demonstrate adaptive control under dynamics that static analytical models cannot easily capture. Adjacent site-specific RAN work further supports realistic wireless-control evaluation: \cite{benzaghta2025data} combines Bayesian optimization, DRL, and Sionna ray tracing for mobility management, while REAL demonstrates closed-loop O-RAN PRB allocation using an OSC near-RT RIC, srsRAN, and a PPO-based xApp under GNU Radio channel impairments~\cite{REAL}. Thus, prior work motivates both joint scheduling--power optimization and realistic closed-loop evaluation, but existing DT/O-RAN studies mainly address BS association, beam selection, slicing, uplink interference, mobility, synchronization, or edge offloading rather than the downlink PRB--power task studied here. Overall, although prior studies provide important building blocks, they do not capture the full combination addressed in this paper: joint \emph{downlink} PRB allocation and power control, a \emph{hierarchical} sequential learning structure, \emph{physically grounded} mobility-aware ray-traced channels, and a \emph{closed-loop} cross-layer execution with HARQ, link adaptation, and matched-channel benchmarking. This combination is important because realistic geometry, mobility, and protocol state fundamentally shape the throughput--fairness tradeoff induced by PRB and power decisions.

\section{System Model and Problem Formulation}
\begin{figure*}[t]
    \centering
    \includegraphics[width=0.66\textwidth]{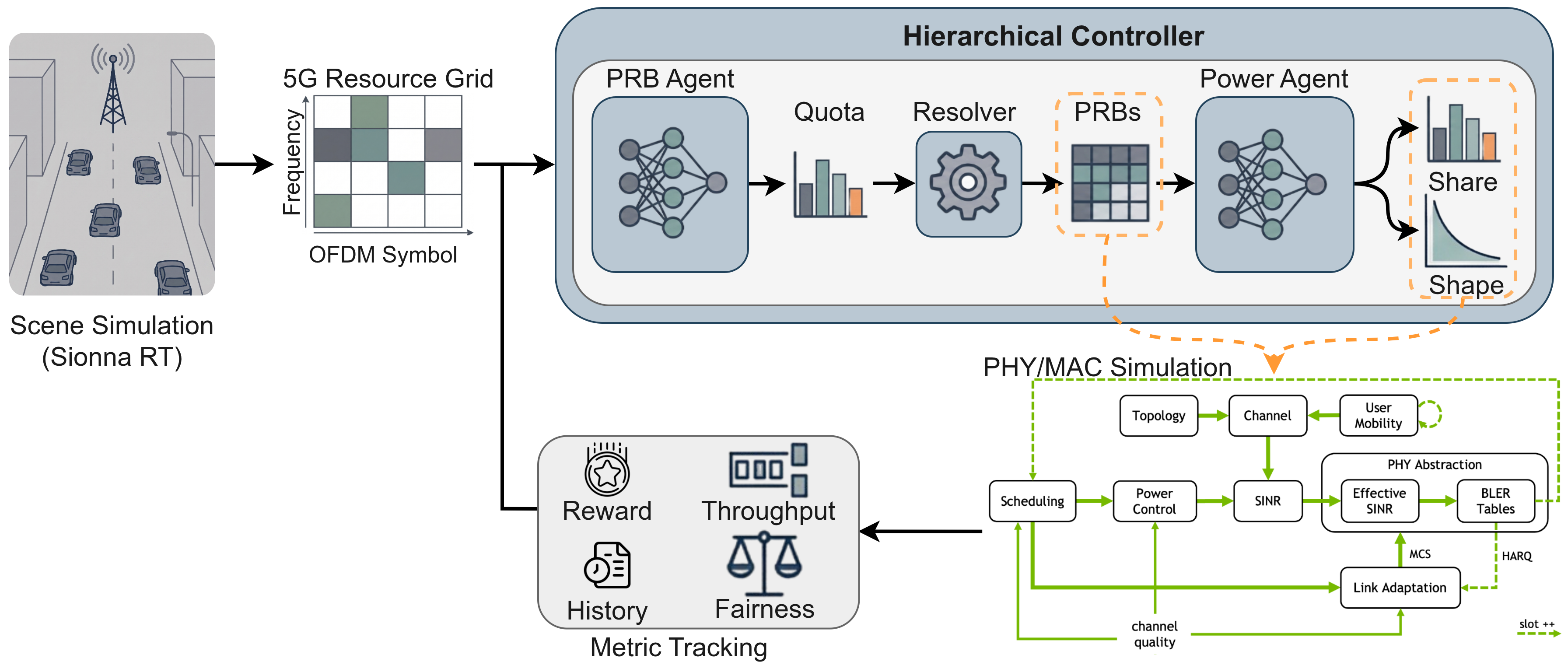}
    \caption{Proposed hierarchical control loop. Sionna RT generates mobility-aware channels. The PRB agent outputs quotas that a deterministic resolver maps to PRBs; the power agent then performs inter-user sharing and intra-user shaping. The PHY/MAC loop with HARQ, OLLA, and MCS adaptation returns throughput/fairness rewards and next observations.}
    \label{fig:system_model}
\end{figure*}
\label{sysModel}

\subsection{Network and Resource Model}

We consider a single-cell downlink OFDMA system in which one base station (BS) serves $U$ mobile user equipments (UEs) over a time-varying wireless channel. The downlink resource grid contains $B$ physical resource blocks (PRBs) per slot and $L$ data OFDM symbols in each slot. Let $t$ denote the slot index, $b \in \{1,\dots,B\}$ the PRB index, $\ell \in \{1,\dots,L\}$ the OFDM symbol index, and $u \in \{1,\dots,U\}$ the user index.

The scheduler determines whether PRB $b$ in slot $t$ is assigned to user $u$ through the binary variable $x_{t,b,u} \in \{0,1\}$, where $x_{t,b,u}=1$ means that PRB $b$ is assigned to user $u$ for the entire slot. Since the downlink is orthogonal across users, each PRB can be assigned to at most one user in a slot:
\begin{equation}
\sum_{u=1}^{U} x_{t,b,u} \leq 1, \quad \forall t,b.
\end{equation}

The same PRB assignment is applied across the $L$ data symbols of the slot, while the transmit power may vary across symbols within an assigned PRB. Let $p_{t,\ell,b,u} \geq 0$ denote the transmit power allocated to user $u$ on PRB $b$ and symbol $\ell$. The BS operates under a total slot-level power constraint:
\begin{equation}
\sum_{\ell=1}^{L}\sum_{b=1}^{B}\sum_{u=1}^{U} p_{t,\ell,b,u} \leq P_{\max}, \quad \forall t.
\end{equation}
To ensure consistency between scheduling and power allocation, power can be allocated only on scheduled PRBs:
\begin{equation}
0 \leq p_{t,\ell,b,u} \leq x_{t,b,u} P_{\max}, \quad \forall t,\ell,b,u.
\end{equation}

\subsection{Sionna System-Level Model and Sionna RT Channel Model}

The overall system-level simulation is implemented using Sionna, while Sionna RT is used to generate scene-aware, mobility-dependent channel realizations. Let $h_{t,\ell,b,u}$ denote the effective channel coefficient for user $u$ on PRB $b$ and symbol $\ell$ at slot $t$. Since the present setting is single-cell and orthogonal, the received signal quality is primarily noise-limited, and the corresponding SINR can be written as
\begin{equation}
\gamma_{t,\ell,b,u} = \frac{p_{t,\ell,b,u} |h_{t,\ell,b,u}|^2}{N_0},
\end{equation}
where $N_0$ is the effective noise power, and a corresponding rate proxy is
\begin{equation}
\hat r_{t,\ell,b,u} = \log_2 \left( 1 + \gamma_{t,\ell,b,u} \right).
\end{equation}
However, the controller does not optimize this proxy alone. Instead, each action is executed through a cross-layer simulation loop in Sionna with adaptive modulation and coding, HARQ feedback, and outer-loop link adaptation (OLLA). Therefore, the delivered throughput depends not only on the instantaneous channel and power allocation, but also on the current state of the link-adaptation process.

\subsection{Fairness-Aware Long-Term Objective}

The goal of this subsection is to define the scalar throughput--fairness objective used to evaluate each slot-level allocation.
Let $R_u(t)$ denote the achieved throughput of user $u$ at slot $t$. To avoid overly myopic behavior, the environment tracks an exponentially smoothed throughput for each user:
\begin{equation}
T_u(t) = (1-\beta) T_u(t-1) + \beta R_u(t),
\end{equation}
where $\beta \in (0,1]$ is the smoothing factor.

To account for both efficiency and service balance, we use Jain's fairness index over the smoothed user throughputs:
\begin{equation}
J_t = \frac{\left(\sum_{u=1}^{U} T_u(t)\right)^2}
{U \sum_{u=1}^{U} T_u^2(t) + \epsilon},
\end{equation}
where $\epsilon > 0$ is a small constant for numerical stability.

The cell-throughput term is normalized as
\begin{equation}
\tilde{T}_t = \mathrm{clip}\left(\frac{\sum_{u=1}^{U} T_u(t)}{T_{\mathrm{norm}}}, 0, 1\right),
\end{equation}
and the instantaneous throughput--fairness objective is defined as
\begin{equation}
g_t = (1-\alpha)\tilde{T}_t + \alpha J_t,
\label{eq:slot_objective}
\end{equation}
where $\alpha \in [0,1]$ controls the throughput--fairness tradeoff.

\subsection{Joint PRB and Power Allocation Problem}

The BS seeks a sequential control rule that jointly determines the slot-level PRB assignment variables $\{x_{t,b,u}\}$ and the symbol-level power-allocation variables $\{p_{t,\ell,b,u}\}$ over time so as to maximize the expected long-term throughput--fairness objective:
\begin{align}
\max_{\mu} \quad & \mathbb{E}_{\mu}\left[\sum_{t=0}^{H-1} \delta^t g_t \right] \\
\text{s.t.} \quad
& x_{t,b,u} \in \{0,1\}, \\
& \sum_{u=1}^{U} x_{t,b,u} \leq 1, \quad \forall t,b, \\
& 0 \leq p_{t,\ell,b,u} \leq x_{t,b,u} P_{\max}, \quad \forall t,\ell,b,u, \\
& \sum_{\ell=1}^{L}\sum_{b=1}^{B}\sum_{u=1}^{U} p_{t,\ell,b,u} \leq P_{\max}, \quad \forall t.
\end{align}
Here, $\delta \in (0,1]$ is the discount factor and $\mu$ denotes a generic sequential control rule, which will be instantiated by the hierarchical MARL controller in the next section.

The formulation above should be viewed as a general sequential control problem. Conventional online optimization is difficult because the integer PRB variables and continuous power variables are coupled through a stateful PHY/MAC loop. A monolithic RL formulation would also be unwieldy: the agent would need to observe channel summaries together with throughput, HARQ/OLLA, and MCS states, while directly outputting a mixed action containing the full PRB map and the symbol-level power tensor. This motivates the hierarchical MARL design in the next section, which separates user-level PRB sharing from conditioned power control while optimizing the same objective.

Figure~\ref{fig:system_model} provides an overview of the end-to-end control loop that connects the Sionna RT scene generation, hierarchical decision making, and cross-layer PHY/MAC execution considered in this work.

\section{Proposed Hierarchical Cooperative MARL Framework}
\label{proposedMethod}

\subsection{Hierarchical Sequential Decision Structure}
Directly learning a joint PRB-and-power allocation policy is difficult because the action space mixes combinatorial scheduling decisions with continuous power control. To address this, we decompose the problem into two cooperative sequential stages within each slot, as illustrated in Fig.~\ref{fig:system_model}. The PRB agent first determines how the available PRBs should be shared across users, and a deterministic resolver then maps these user-level quotas into exact PRB assignments. Conditioned on the resolved schedule, the power agent allocates the BS transmit-power budget across users and shapes that power over the assigned resources. The resulting allocation is executed in the Sionna-based PHY/MAC loop, whose throughput, fairness, HARQ, and link-adaptation outcomes are fed back to the controller.

We refer to the framework as cooperative MARL because two distinct policies, with different observations and action spaces, act sequentially within each slot and are trained to optimize the same long-term objective. In the MARL implementation, the slot-level objective in \eqref{eq:slot_objective} is used as the common reward, i.e., $r_t \triangleq g_t$. Although execution is staged, the two agents cooperate through the shared environment and common reward.

Formally, let $s_t^{\mathrm{prb}}$ and $s_t^{\mathrm{pow}}$ denote the observations of the PRB and power agents at slot $t$. The control process follows
\begin{equation}
s_t^{\mathrm{prb}} \rightarrow a_t^{\mathrm{prb}} \rightarrow \mathbf{X}_t
\rightarrow s_t^{\mathrm{pow}} \rightarrow a_t^{\mathrm{pow}} \rightarrow \mathbf{P}_t,
\end{equation}
where $\mathbf{X}_t$ is the PRB assignment and $\mathbf{P}_t$ is the resulting power-allocation tensor.

\subsection{PRB Allocation via Quota Learning and Deterministic Resolution}

Rather than outputting a full combinatorial PRB map, the PRB agent outputs a compact user-level vector
\begin{equation}
\mathbf{z}_t=[z_{t,1},\dots,z_{t,U}],
\end{equation}
which is converted into normalized PRB shares and integer quotas as
% \vspace{-4mm}
\begin{equation}
q_{t,u}=\frac{e^{z_{t,u}}}{\sum_{v=1}^{U}e^{z_{t,v}}},
\qquad
\bar{B}_{t,u}=q_{t,u}B,
\qquad
\sum_{u=1}^{U}B_{t,u}=B.
\end{equation}
The integer quotas $B_{t,u}$ are obtained by flooring $\bar{B}_{t,u}$ and applying largest-remainder correction.

Given these quotas, a deterministic channel-aware resolver assigns exact PRB indices. Since power is selected only after scheduling, the resolver ranks PRBs using the channel-only score
\begin{equation}
\psi_{t,b,u}=\sum_{\ell=1}^{L}|h_{t,\ell,b,u}|^2 .
\end{equation}
Let $\mathcal{B}^{\mathrm{avail}}_t$ be the set of unassigned PRBs and
$\mathcal{U}^{\mathrm{act}}_t=\{u:B_{t,u}>0\}$ the users with remaining quota. The resolver cycles over active users and assigns each user its best available PRB,
\begin{equation}
b_t^\star(u)=\arg\max_{b\in\mathcal{B}^{\mathrm{avail}}_t}\psi_{t,b,u},
\qquad
x_{t,b_t^\star(u),u}=1,
\end{equation}
then removes $b_t^\star(u)$ from $\mathcal{B}^{\mathrm{avail}}_t$ and decrements $B_{t,u}$ until all quotas are exhausted. The resulting PRB map is reused across all $L$ data symbols, while symbol-level powers are determined by the power agent. Thus, the policy learns \emph{how much} spectrum each user receives, and the resolver determines \emph{which} PRBs are assigned, guaranteeing feasibility without learning a full PRB map.

\subsection{Factorized Power Allocation}

After the PRB allocation is determined, the power agent allocates the BS power budget in a structured way. Its action is factorized into a user-level weight vector $\mathbf{w}_t = [w_{t,1},\dots,w_{t,U}]$ and a shaping-coefficient vector $\boldsymbol{\kappa}_t=[\kappa_{t,1},\dots,\kappa_{t,U}]$.

The user-level weights are first converted into normalized power shares:
\begin{equation}
\eta_{t,u} = \frac{e^{w_{t,u}}}{\sum_{v=1}^{U} e^{w_{t,v}}},
\end{equation}
which define the per-user power budgets
\begin{equation}
P^{\mathrm{tot}}_{t,u} = \eta_{t,u} P_{\max},
\end{equation}
so that $\sum_{u=1}^{U} P^{\mathrm{tot}}_{t,u}=P_{\max}$.

The second component controls how each user's power budget is distributed across its scheduled PRB-symbol resources. Let
\begin{equation}
\mathcal{B}_{t,u}=\{b:x_{t,b,u}=1\}
\end{equation}
denote the set of PRBs assigned to user $u$ in slot $t$, and let $\rho_{t,\ell,u}(m)$ denote the PRB with rank $m$ after sorting $\mathcal{B}_{t,u}$ in descending order of $|h_{t,\ell,b,u}|^2$ for symbol $\ell$.
We then define the exponential shaping weights as
\begin{equation}
\omega_{t,\ell,\rho_{t,\ell,u}(m),u}
=
\exp\!\left(-\kappa_{t,u}(m-1)\right),
\end{equation}
for $m=1,\dots,|\mathcal{B}_{t,\ell,u}|$. The resulting power allocated to user $u$ on symbol $\ell$ and PRB $b$ is

\begin{equation}
p_{t,\ell,b,u}
=
x_{t,b,u}\,P^{\mathrm{tot}}_{t,u}
\frac{\omega_{t,\ell,b,u}}
{\sum_{\ell'=1}^{L}\sum_{b'=1}^{B} x_{t,b',u}\,\omega_{t,\ell',b',u}}.
\end{equation}

When $\kappa_{t,u}\approx 0$, the power budget of user $u$ is distributed nearly uniformly over its scheduled PRB-symbol resources, whereas larger values of $\kappa_{t,u}$ increasingly concentrate power on the stronger channel-ranked resources. The resulting symbol-by-PRB power tensor is finally expanded uniformly over the $12$ subcarriers within each scheduled PRB to obtain the RE-level power tensor used by the PHY abstraction.

\subsection{Observation Design and Curriculum Training}

%Both agents use compact cross-layer observations rather than raw channel tensors. Each user's feature vector includes channel-quality summaries together with historical service and reliability indicators, such as previous allocation share, smoothed throughput, HARQ success history, and MCS state. The PRB agent observes these features before scheduling, while the power agent observes them after the PRB assignment has been fixed.
%% COMPRESSED VERSION
Both agents use compact cross-layer observations rather than raw channel tensors. Per-user features include channel-quality summaries, previous allocation share, smoothed throughput, HARQ success history, and MCS state. The PRB agent observes them before scheduling; the power agent observes them after PRB resolution.

Both agents are trained using PPO. Since joint exploration over scheduling and power control can be unstable, we adopt a staged curriculum-based training procedure. In the first phase, only the PRB agent is trained while power is assigned using a simple baseline rule. In the second phase, the power agent is trained on top of the learned PRB behavior. In the final phase, both agents are fine-tuned jointly. This strategy matches the hierarchy of the control problem and improves training stability in the Sionna-based environment. The intermediate PRB-only policy obtained after the first phase is also used as an ablation in evaluation, while a separate power-only ablation is trained by fixing the PRB scheduler to PF and learning only the power policy.

% Overall, the proposed framework combines learning-based control with domain-structured action resolution. The PRB stage reduces combinatorial complexity through quota learning, the power stage captures fine-grained energy allocation through a factorized design, and the staged curriculum improves optimization stability in a physically grounded cross-layer setting.

\begin{algorithm}[t]
\caption{Training and Inference of the Proposed Hierarchical Controller}
\label{alg:hmarl}
\begin{algorithmic}[1]
\STATE Initialize PRB policy $\pi_{\mathrm{prb}}$ and power policy $\pi_{\mathrm{pow}}$
\FOR{phase $=1,2,3$}
    \IF{phase $=1$}
        \STATE Train $\pi_{\mathrm{prb}}$ only; use equal-power transmission
    \ELSIF{phase $=2$}
        \STATE Freeze $\pi_{\mathrm{prb}}$ and train $\pi_{\mathrm{pow}}$
    \ELSE
        \STATE Fine-tune $\pi_{\mathrm{prb}}$ and $\pi_{\mathrm{pow}}$ jointly
    \ENDIF
    \FOR{each PPO iteration}
        \STATE Reset the environment
        \FOR{each slot $t$}
            \STATE Observe $s_t^{\mathrm{prb}}$ and sample $a_t^{\mathrm{prb}}$
            \STATE Convert $a_t^{\mathrm{prb}}$ into quotas $\{B_{t,u}\}_{u=1}^{U}$
            \STATE Resolve exact PRB assignment $\mathbf{X}_t$ using the deterministic channel-aware resolver
            \IF{phase $=1$}
                \STATE Apply equal-power allocation and execute the PHY step
            \ELSE
                \STATE Construct $s_t^{\mathrm{pow}}$ from the resolved schedule
                \STATE Observe $s_t^{\mathrm{pow}}$ and sample $a_t^{\mathrm{pow}}$
                \STATE Compute inter-user power shares and intra-user exponential shaping
                \STATE Build the RE-level power tensor $\mathbf{P}_t$ and execute the PHY step
            \ENDIF
            \STATE Receive common reward $r_t$ and update throughput, HARQ, and MCS trackers
        \ENDFOR
        \STATE Update the trainable policy or policies using PPO
    \ENDFOR
\ENDFOR
\STATE \textbf{Inference:} For each slot, apply $\pi_{\mathrm{prb}}$, deterministic PRB resolution, $\pi_{\mathrm{pow}}$, and the PHY execution without exploration
\end{algorithmic}
\end{algorithm}
\section{Simulation Setup and Evaluation Methodology}
\label{simulation}
\begin{figure}[t]
    \centering
    \subfloat[Ray-tracing scene]{%
        \includegraphics[width=0.48\columnwidth]{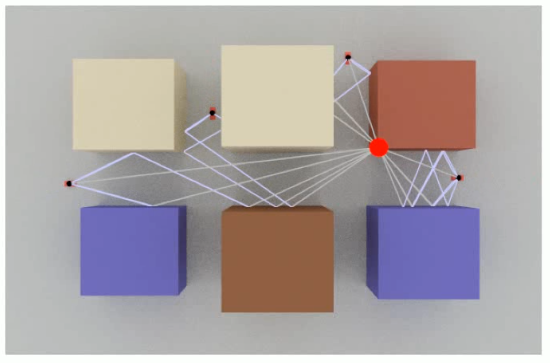}
    }\hfill
    \subfloat[Radio map]{%
        \includegraphics[width=0.48\columnwidth]{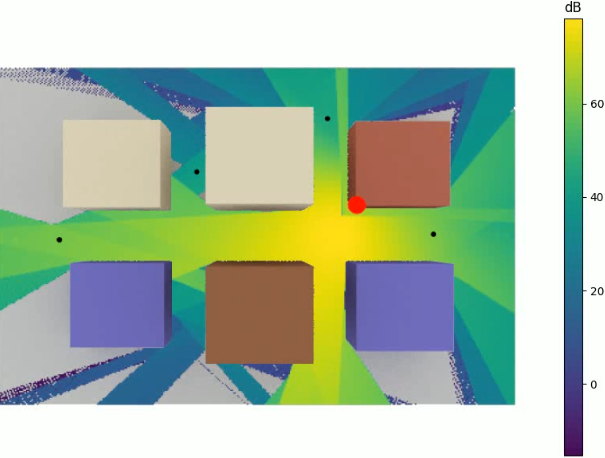}
    }
    \caption{Sionna RT evaluation environment. (a) Ray-tracing visualization of the urban scene with a rooftop BS (red) and four vehicle-mounted receivers (black). (b) Corresponding radio map showing the spatial distribution of received power over the same environment.}
    \label{fig:env_radio}
\end{figure}

\subsection{Sionna-Based Simulation Environment}

We evaluate a physically grounded single-cell 5G downlink using Sionna for system-level simulation and Sionna RT for scene construction, mobility-aware propagation, and ray-traced channels. Fig.~\ref{fig:env_radio} shows a rooftop BS serving four vehicle-mounted receivers in a street-canyon environment whose radio map captures geometry-dependent received-power variation.

%The system operates at a carrier frequency of $3.5$ GHz with $30$ kHz subcarrier spacing and $51$ PRBs. Each PRB contains $12$ subcarriers, and each slot uses $12$ OFDM symbols for data transmission; the remaining two symbols of the nominal $14$-symbol slot are reserved for signaling/control, the BS transmit power is $40$ dBm, the UE noise figure is $7$ dB, and the slot duration is $0.5$ ms. User mobility is explicitly modeled through predefined trajectories and different speeds, so the wireless channel evolves with scene geometry rather than with a purely synthetic fading process.
%% COMPRESSED VERSION
The carrier frequency is $3.5$ GHz with $30$ kHz subcarrier spacing, $51$ PRBs, $12$ subcarriers per PRB, and $12$ data OFDM symbols per $14$-symbol slot; two symbols are reserved for signaling/control. The BS power is $40$ dBm, the UE noise figure is $7$ dB, and the slot duration is $0.5$ ms. User mobility follows predefined trajectories and speeds, producing geometry-driven channel evolution rather than purely synthetic fading.

% \vspace{-2mm}
\subsection{Cross-Layer Execution}
% \vspace{-2mm}
At every slot, the environment first obtains the current channel realization from Sionna RT. The PRB agent then selects user-level resource shares, which are converted into exact PRB assignments through the deterministic quota-based resolver. Conditioned on this schedule, the power agent determines how the total BS power budget is distributed across users and across their assigned REs. This interaction is summarized in Fig.~\ref{fig:system_model}, which highlights the sequential coupling between scene-aware channel generation, hierarchical control, and PHY/MAC feedback.

The resulting scheduling and power tensors are then passed to the Sionna-based system-level simulation loop, which performs effective SINR computation, adaptive MCS selection, HARQ feedback, and outer-loop link adaptation. The environment updates the throughput, reliability, and allocation-history trackers used to construct the next state and common reward. This ensures that the learned controller is evaluated in a closed-loop wireless system rather than through a static rate model.
% \vspace{-2mm}
\subsection{Comparison Schemes and Metrics}
We compare four schemes under the same evaluation protocol. The first is the PF baseline with equal-power transmission, denoted PF (Baseline). For each PRB $b$ in slot $t$, the PF scheduler selects
% \vspace{-4mm}
\begin{equation}
u_{t,b}^{\star}
=
\arg\max_{u}
\frac{\psi_{t,b,u}}{T_u(t)+\epsilon},
\end{equation}
where $\psi_{t,b,u}=\sum_{\ell=1}^{L} |h_{t,\ell,b,u}|^2$ is the slot-level channel-quality score and $T_u(t)$ is the smoothed throughput defined earlier. Thus, PF favors users with strong instantaneous channels while preventing persistent domination by already well-served users. Power is then distributed uniformly over the scheduled PRB-symbol resources. 
To isolate the contribution of each learned component, we also evaluate two ablation variants. The PRB Agent uses the PRB policy trained in the first curriculum phase and applies uniform power over the scheduled resources. The Power Agent uses the PF scheduler for PRB assignment and applies a separately trained power policy conditioned on the PF schedule. Finally, PRB+Power denotes the proposed hierarchical controller after joint fine-tuning of both agents. 

Performance is evaluated using empirical CDFs and summary statistics of cell throughput and Jain's fairness index under the matched-channel protocol. We report the mean, median, and 10th-percentile cell throughput, together with the mean and median Jain's fairness index, so that both average behavior and lower-tail performance can be compared across schemes.

\subsection{Matched-Channel Evaluation Protocol}

To ensure a fair comparison, all comparison schemes are evaluated on the same cached sequence of Sionna RT channel realizations. Before evaluation run, the internal PHY/MAC states of all systems are reset so that HARQ memory, link-adaptation state, and related historical variables start from identical conditions.

This matched-channel protocol is important because it removes channel randomness as a source of performance variation. Any observed difference among the compared schemes can therefore be attributed more directly to the quality of the control policy itself. The same sequential execution logic used during training is preserved during evaluation, with the PRB decision applied first and the power decision applied second.

\section{Results and Discussion}
\label{results}

\begin{figure}[t]
    \centering
    \subfloat[Cell throughput]{%
        \includegraphics[width=0.48\columnwidth]{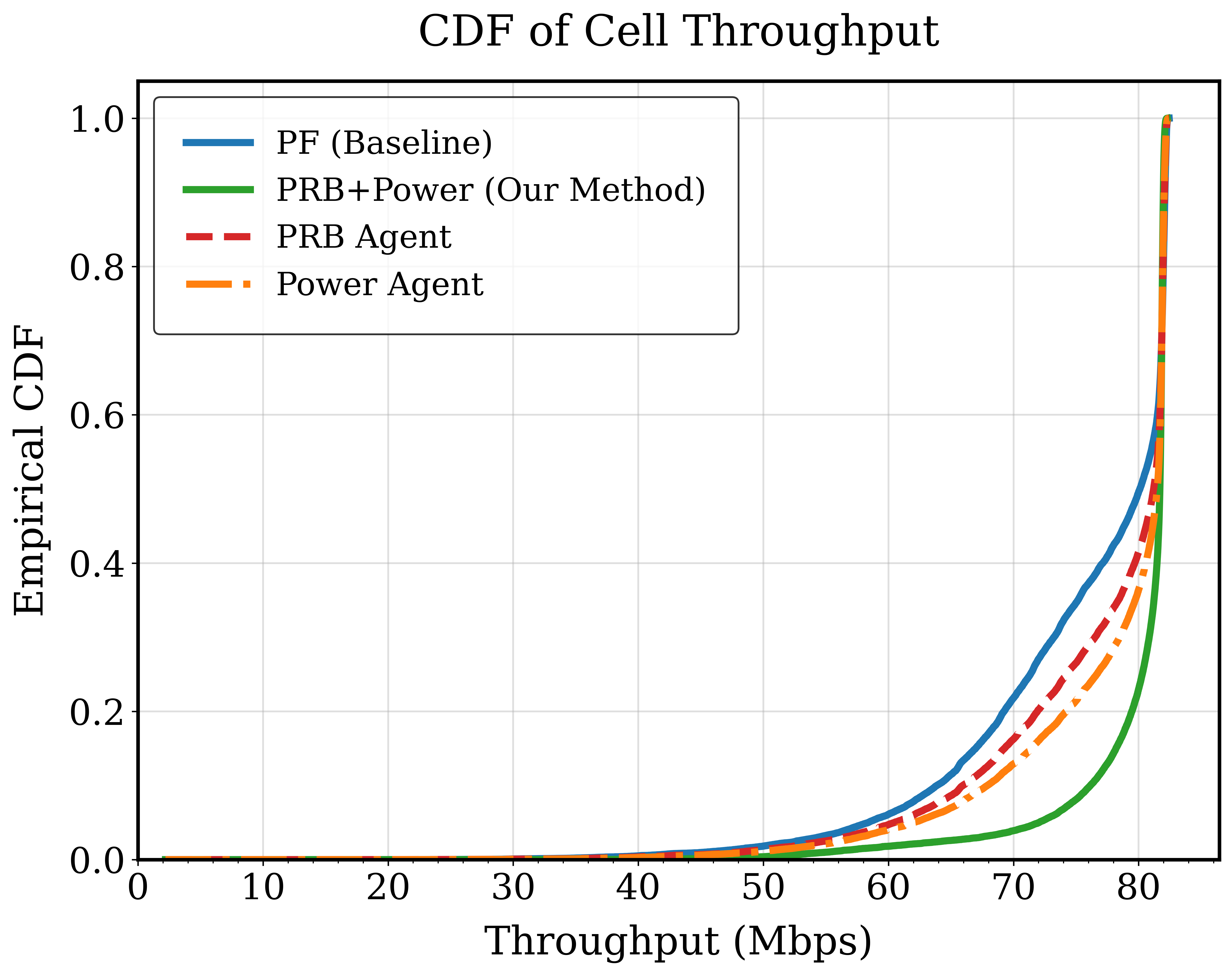}
        \label{fig:cdf_thr}
    }\hfill
    \subfloat[Jain's fairness index]{%
        \includegraphics[width=0.48\columnwidth]{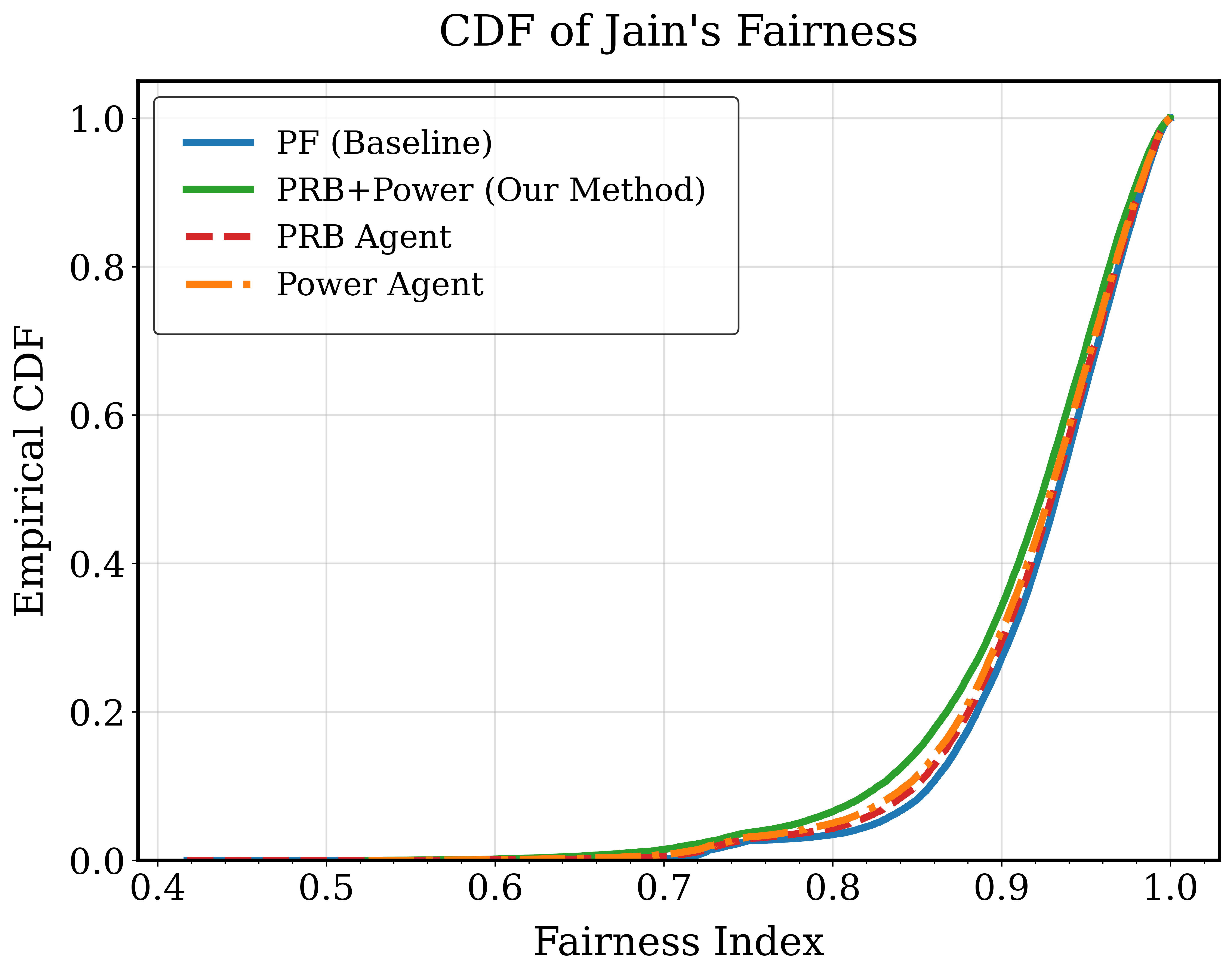}
        \label{fig:cdf_fair}
    }
    \caption{Empirical CDFs under the matched-channel evaluation protocol: (a) cell throughput and (b) Jain's fairness index. The PRB Agent and Power Agent curves isolate the effects of learned scheduling and learned power allocation, respectively, while PRB+Power denotes the proposed joint controller.}
    \label{fig:cdf_results}
\end{figure}

\begin{table}[t]
\centering
\caption{Quantitative comparison under matched-channel evaluation. Throughput is reported in Mbps, and $\Delta\bar{T}$ is the mean throughput gain relative to PF.}
\label{tab:quant_results}
\footnotesize
\setlength{\tabcolsep}{2.5pt}
\resizebox{\columnwidth}{!}{
\begin{tabular}{lcccccc}
\hline
Scheme & $\bar{T}$ & $\Delta\bar{T}$ & $T_{10}$ & $T_{50}$ & $\bar{J}$ & $J_{50}$ \\
\hline
PF (Baseline) & 75.67 & -- & 63.87 & 80.10 & 0.9233 & 0.9337 \\
PRB Agent & 76.98 & +1.73\% & 65.94 & 81.20 & 0.9193 & 0.9311 \\
Power Agent & 77.77 & +2.77\% & 67.89 & 81.48 & 0.9168 & 0.9294 \\
PRB+Power & 79.90 & +5.58\% & 76.14 & 81.70 & 0.9101 & 0.9248 \\
\hline
\end{tabular}}
\end{table}

%Fig.~\ref{fig:cdf_results} compares the PF equal-power baseline, the two learned ablations, and the proposed PRB+Power controller under the matched-channel evaluation protocol. In Fig.~\ref{fig:cdf_thr}, both the PRB Agent and Power Agent shift the cell-throughput CDF to the right of the PF baseline, showing that learned scheduling and learned power control each provide measurable gains when applied independently. The proposed PRB+Power controller gives the largest right shift and the steepest high-throughput behavior, indicating that the two learned components are complementary rather than redundant.

Fig.~\ref{fig:cdf_results} and Table~\ref{tab:quant_results} compare the PF equal-power baseline, the two learned ablations, and the proposed PRB+Power controller under the matched-channel evaluation protocol. Both ablations improve the throughput distribution relative to PF: the PRB Agent increases the mean cell throughput by $1.73\%$, while the Power Agent increases it by $2.77\%$. The full PRB+Power controller achieves the largest gain, improving the mean cell throughput from $75.67$ Mbps to $79.90$ Mbps, corresponding to a $5.58\%$ improvement over PF.

The Jain's fairness CDF in Fig.~\ref{fig:cdf_fair} shows the expected throughput--fairness tradeoff. PF maintains the strongest fairness behavior, while the learned variants slightly shift the distribution toward lower fairness values. However, the degradation remains modest relative to the throughput improvement, especially for the full PRB+Power controller. This indicates that the proposed method improves cell efficiency primarily by exploiting favorable PRB and power-allocation opportunities, while still preserving a high level of long-term service balance. 

The lower-tail throughput also improves substantially. The 10th-percentile cell throughput increases from $63.87$ Mbps under PF to $76.14$ Mbps with PRB+Power, giving a $19.21\%$ gain. This indicates that the proposed controller improves not only average cell efficiency, but also the lower-throughput operating region. The Jain's fairness results show the expected throughput--fairness tradeoff: the median fairness index decreases from $0.9337$ under PF to $0.9248$ with PRB+Power, which is a relative reduction of only $0.95\%$.

%Overall, the ablation results strengthen the main claim of the paper. The PRB-only and power-only variants each outperform PF in throughput, but the full hierarchical controller achieves the best throughput distribution by jointly adapting resource shares and power shaping in the Sionna RT-based cross-layer environment.
Overall, the ablation results strengthen the main claim of the paper. The PRB-only and power-only variants each outperform PF in throughput, confirming that learned scheduling and learned power allocation are individually useful. Their combination achieves the best throughput distribution, especially in the lower tail, while preserving a high fairness level in the considered Sionna RT-based 5G environment.
% \vspace{-2mm}
\section{Conclusion}
% \vspace{-2mm}
\label{conclusion}
%This paper presented a hierarchical cooperative MARL framework for joint downlink PRB and power allocation in a physically grounded 5G environment built with Sionna and Sionna RT. The proposed design combines quota-based PRB allocation, deterministic channel-aware PRB resolution, and factorized power control within a cross-layer loop that includes HARQ and link adaptation. Under matched-channel evaluation against PF equal-power scheduling and two learned ablations, the proposed PRB+Power controller achieves the strongest improvement in cell throughput while incurring only a modest reduction in Jain's fairness index. These results show that the hierarchical decomposition is effective in translating coordinated scheduling and power adaptation into a better throughput--fairness operating point in realistic ray-traced wireless channels. Future work will extend the framework to multi-cell scenarios, queue-aware traffic models, and more advanced multi-antenna settings.
%% COMPRESSED VERSION
This paper presented a hierarchical cooperative MARL framework for joint downlink PRB and power allocation in a Sionna/Sionna RT 5G environment. The design combines quota-based PRB allocation, deterministic channel-aware resolution, and factorized power control within a HARQ/link-adaptation loop. Under matched-channel evaluation against PF equal-power scheduling and two ablations, PRB+Power delivers the strongest throughput improvement with only modest Jain-fairness reduction, showing that hierarchical decomposition can translate coordinated scheduling and power adaptation into a better throughput--fairness operating point. Future work will extend the framework to multi-cell, queue-aware, and advanced multi-antenna settings.
% \vspace{-3mm}
\bibliographystyle{IEEEtran}
\bibliography{bibtex}

@INPROCEEDINGS{REAL,
  author={Barker, Ryan and Dorcheh, Alireza Ebrahimi and Seyfi, Tolunay and Afghah, Fatemeh},
  booktitle={2025 IEEE International Conference on Communications Workshops (ICC Workshops)}, 
  title={REAL: Reinforcement Learning-Enabled xApps for Experimental Closed-Loop Optimization in O-RAN with OSC RIC and srsRAN}, 
  year={2025},
  volume={},
  number={},
  pages={389-395},
  doi={10.1109/ICCWorkshops67674.2025.11162144}}

@INPROCEEDINGS{DORA,
  author={Dorcheh, Alireza Ebrahimi and Seyfi, Tolunay and Afghah, Fatemeh},
  booktitle={2025 IEEE Middle East Conference on Communications and Networking (MECOM)}, 
  title={DORA: Dynamic O-RAN Resource Allocation for Multi-Slice 5G Networks}, 
  year={2025},
  volume={},
  number={},
  pages={1-6},
  doi={10.1109/MECOM67453.2025.11439664}}

@techreport{3gpp38211,
  author      = {{3GPP}},
  title       = {{NR; Physical Channels and Modulation}},
  institution = {{3rd Generation Partnership Project (3GPP)}},
  type        = {{Technical Specification}},
  number      = {{TS 38.211}},
  year        = {2025},
  note        = {{Release 18, Version 18.7.0}}
}

@techreport{3gpp38214,
  author      = {3GPP},
  title       = {{NR; Physical Layer Procedures for Data}},
  institution = {{3rd Generation Partnership Project (3GPP)}},
  type        = {{Technical Specification}},
  number      = {{TS 38.214}},
  year        = {2025},
  note        = {{Release 17, Version 17.14.0}}
}

@techreport{3gpp38321,
  author      = {3GPP},
  title       = {{NR; Medium Access Control (MAC) Protocol Specification}},
  institution = {{3rd Generation Partnership Project (3GPP)}},
  type        = {{Technical Specification}},
  number      = {{TS 38.321}},
  year        = {2025},
  note        = {{Release 18}}
}

@techreport{3gpp38104,
  author      = {{3GPP}},
  title       = {{NR; Base Station (BS) Radio Transmission and Reception}},
  institution = {{3rd Generation Partnership Project (3GPP)}},
  type        = {{Technical Specification}},
  number      = {{TS 38.104}},
  year        = {2025},
  note        = {{Release 18, Version 18.10.0}}
}

@article{kelly1998rate,
  author  = {Kelly, Frank P. and Maulloo, Aman K. and Tan, David K. H.},
  title   = {Rate Control for Communication Networks: Shadow Prices, Proportional Fairness and Stability},
  journal = {Journal of the Operational Research Society},
  volume  = {49},
  number  = {3},
  pages   = {237--252},
  year    = {1998}
}

@inproceedings{jalali2000data,
  author    = {Jalali, Ahmad and Padovani, Roberto and Pankaj, Ramesh},
  title     = {Data Throughput of CDMA-HDR: A High Efficiency-High Data Rate Personal Communication Wireless System},
  booktitle = {Proc. IEEE Vehicular Technology Conference (VTC)},
  pages     = {1854--1858},
  year      = {2000}
}

@article{kushner2004convergence,
  author  = {Kushner, Harold J. and Whiting, Philip A.},
  title   = {Convergence of Proportional-Fair Sharing Algorithms under General Conditions},
  journal = {IEEE Transactions on Wireless Communications},
  volume  = {3},
  number  = {4},
  pages   = {1250--1259},
  year    = {2004}
}

@article{jang2025joint,
  author       = {Jonggyu Jang and Hyeonsu Lyu and David J. Love and Hyun Jong Yang},
  title        = {Joint Optimization of User Association and Resource Allocation for Load Balancing With Multi-Level Fairness},
  journal      = {arXiv preprint arXiv:2505.08573},
  year         = {2025},
  eprint       = {2505.08573},
  archivePrefix= {arXiv},
  primaryClass = {eess.SP}
}

@article{tarafder2026digital,
  author       = {Pulok Tarafder and Zoheb Hassan and Imtiaz Ahmed and Danda B. Rawat and Kamrul Hasan and Cong Pu},
  title        = {Digital-Twin Empowered Deep Reinforcement Learning For Site-Specific Radio Resource Management in NextG Wireless Aerial Corridor},
  journal      = {arXiv preprint arXiv:2602.03801},
  year         = {2026},
  eprint       = {2602.03801},
  archivePrefix= {arXiv},
  primaryClass = {eess.SP},
  note         = {Submitted for possible publication to IEEE}
}

@article{elloumi2026uplink,
  author       = {Mohamed Elloumi and Md. Zoheb Hassan and Georges Kaddoum},
  title        = {Uplink Radio Resource Block and Power Coordination in Open RAN-Digital Twin-Integrated Multi-Cell Internet of Drone Networks},
  journal      = {TechRxiv},
  year         = {2026},
  month        = feb,
  doi          = {10.36227/techrxiv.177004951.16796998/v1},
  note         = {Posted on 2 Feb 2026; preprint}
}

@inproceedings{chen2023digital,
  author       = {Xiangchun Chen and Jiannong Cao and Zhixuan Liang and Yuvraj Sahni and Mingjin Zhang},
  title        = {Digital Twin-assisted Reinforcement Learning for Resource-aware Microservice Offloading in Edge Computing},
  booktitle    = {2023 IEEE 20th International Conference on Mobile Ad Hoc and Smart Systems (MASS)},
  year         = {2023},
  doi          = {10.1109/MASS58611.2023.00012}
}

@article{tong2025continual,
  author       = {Haonan Tong and Mingzhe Chen and Jun Zhao and Ye Hu and Zhaohui Yang and Yuchen Liu and Changchuan Yin},
  title        = {Continual Reinforcement Learning for Digital Twin Synchronization Optimization},
  journal      = {IEEE Transactions on Mobile Computing},
  volume       = {24},
  number       = {8},
  pages        = {6843--6857},
  year         = {2025},
  doi          = {10.1109/TMC.2025.3546507}
}

@article{benzaghta2025data,
  author       = {Mohamed Benzaghta and Sahar Ammar and David L{\'o}pez-P{\'e}rez and Basem Shihada and Giovanni Geraci},
  title        = {Data-Driven Cellular Mobility Management via Bayesian Optimization and Reinforcement Learning},
  journal      = {arXiv preprint arXiv:2505.21249},
  year         = {2025},
  eprint       = {2505.21249},
  archivePrefix= {arXiv},
  primaryClass = {cs.IT}
}

@inproceedings{setayesh2020joint,
  author    = {Mehdi Setayesh and Shahab Bahrami and Vincent W. S. Wong},
  title     = {Joint {PRB} and Power Allocation for Slicing {eMBB} and {URLLC} Services in {5G} {C-RAN}},
  booktitle = {Proc. IEEE Global Communications Conference (GLOBECOM)},
  year      = {2020}
}

@inproceedings{elsayed2019reinforcement,
  author    = {Medhat Elsayed and Melike Erol{-}Kantarci},
  title     = {Reinforcement Learning{-}Based Joint Power and Resource Allocation for {URLLC} in {5G}},
  booktitle = {Proc. IEEE Global Communications Conference (GLOBECOM)},
  year      = {2019},
  doi       = {10.1109/GLOBECOM38437.2019.9014032}
}

@inproceedings{cheng2023joint,
  author    = {Qian Cheng and Kang Li and Pengcheng Zhu and Jiamin Li and Yanxiang Jiang and Dongming Wang},
  title     = {Joint Resource Block and Power Allocation for {eMBB} and {URLLC} Coexistence in {5G} {H-CRAN}},
  booktitle = {Proc. International Conference on Wireless Communications and Signal Processing (WCSP)},
  year      = {2023},
  doi       = {10.1109/WCSP58612.2023.10405139}
}

@inproceedings{li2023joint,
  author    = {Xiaodong Li and Weixi Zhou and Hongjie Zhang and Jing Zhao and Dongcai Zhao and Zhicheng Dong},
  title     = {Joint Subcarrier and Power Allocation in Mobile Scenario of the {OFDM} Systems Based on Deep Reinforcement Learning},
  booktitle = {Proc. International Conference on Computer, Communication and Control Systems (ICCCS)},
  pages     = {209--214},
  year      = {2023}
}

@article{kim2025joint,
  author    = {Joeun Kim and Youngil Jeon and Junhwan Lee and Moon{-}Sik Lee and Taesoo Kwon},
  title     = {Joint Scheduling and Resource Allocation Based on Reinforcement Learning in Integrated Access and Backhaul Networks},
  journal   = {ICT Express},
  volume    = {11},
  number    = {3},
  pages     = {536--541},
  year      = {2025},
  doi       = {10.1016/j.icte.2025.03.004}
}

\end{document}